\documentclass[10pt,letterpaper]{article}
\usepackage[top=0.85in,left=1.5in,footskip=0.75in,marginparwidth=2in]{geometry}
\usepackage{amsmath}

\usepackage[utf8]{inputenc} 
\usepackage[T1]{fontenc} 
\usepackage{lmodern} 
\usepackage{graphicx}
\usepackage{subcaption}
\usepackage{mathtools}
\usepackage{verbatim}
\usepackage{amssymb}
\usepackage{gensymb}
\usepackage{upgreek}

\usepackage{cite}

\usepackage{nameref,hyperref}

\usepackage{microtype}
\DisableLigatures[f]{encoding = *, family = * }

\usepackage{changepage}

\usepackage[aboveskip=1pt,labelfont=bf,labelsep=period,singlelinecheck=off]{caption}

\usepackage{lastpage,fancyhdr,graphicx}
\usepackage{epstopdf}
\fancyheadoffset[L]{2.25in}
\fancyfootoffset[L]{2.25in}

\usepackage{color}

\definecolor{Gray}{gray}{.25}

\usepackage{graphicx}

\begin{document}
\vspace*{0.35in}

\begin{flushleft}
{\LARGE
\textbf\newline{Suppression of Kerr-Induced Satellites in Multi-Pulse CPA}
}
\newline

\textbf{Vinzenz Stummer\textsuperscript{1,*},
Edgar Kaksis\textsuperscript{1},
Audrius Pugžlys\textsuperscript{1,2},
Andrius Baltuška\textsuperscript{1,2}
}
\\
\bigskip
[1] Photonics Institute, TU Wien, Gusshausstrasse 27/387, 1040 Vienna, Austria
\newline
[2] Center for Physical Sciences \& Technology, Savanoriu Ave. 231 LT-02300 Vilnius, Lithuania
\\
\bigskip
* vinzenz.stummer@tuwien.ac.at

\end{flushleft}

\section*{Abstract}
Amplification of bursts of ultrashort pulses is very challenging when the intraburst repetition frequency reaches the THz range, corresponding to (sub)-ps intervals between consecutive pulses. Periodic interference significantly modifies conditions for chirped pulse amplification (CPA) which gives rise to temporal and spectral distortions during CPA due to the optical Kerr nonlinearity. Multi-pulse chirped amplification to mJ energies may lead to a pronounced degradation of burst fidelity and the appearance of periodic temporal satellites after de-chirping the amplified waveform. We study, experimentally and numerically, the limitations of THz burst-mode CPA caused by self- and cross-phase modulation. A number of practical recipes to suppress nonlinear distortions and improve energy scaling by optimizing burst parameters and applying modulation techniques are presented.

\section{Introduction}
Chirped Pulse Amplification (CPA) allows the generation of high-energy ultrashort laser pulses \cite{strickland-cpa}, up to multiple hundreds of Joules \cite{PW-laser_review}. However, a problematic case is given in CPA when chirped pulses are amplified at once in a single amplifier with a chirped pulse duration $\tau_s$ that is much larger than their temporal spacing $\Delta t$ ($\tau_s \gg \Delta t$). If the burst pulses are strongly stretched, the pulse overlap and interference effects determine the chirped waveform. The result is a sequence of intensity peaks centred around wavelengths $\lambda_i$. For sub-millijoule amplification or higher, each $\lambda_i$-maximum experiences Kerr-induced Self-Phase Modulation (SPM) within the amplifier. The consequence is the generation of $\Delta t$-periodic satellite pulses in CPA \cite{didenko2008}. 

\begin{figure}[htbp]
    \centering
    \includegraphics[width=\textwidth]{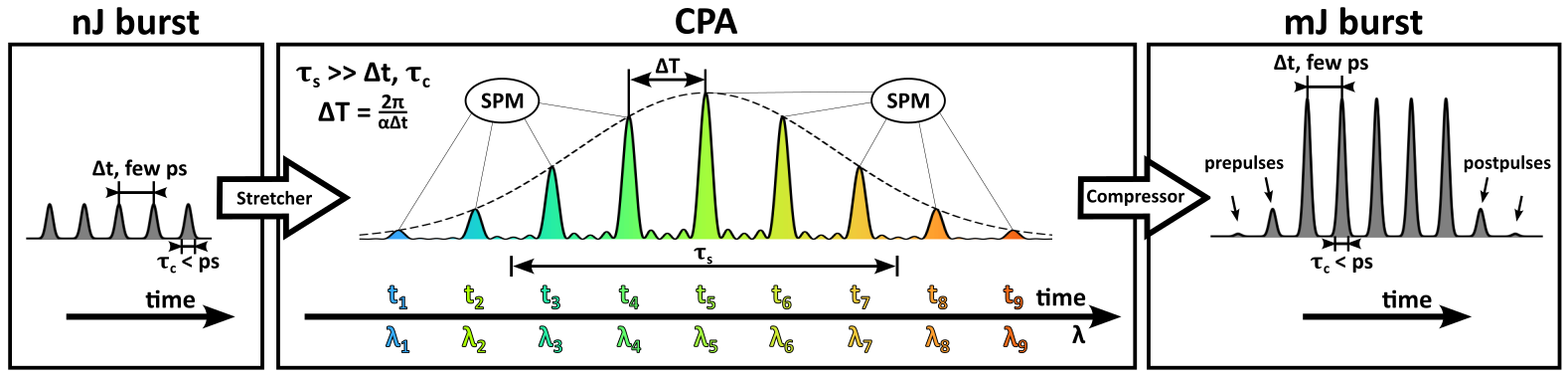}
    \caption{\textbf{Satellite formation in multi-pulse Chirped-Pulse Amplification (CPA).} When stretching ultrashort pulses with sub-picosecond pulse duration $\tau_c$ and few-picosecond spacing $\Delta t$ to a sufficiently long chirped pulse duration $\tau_s$ with a chirp rate $d\omega(t)/dt = \alpha$, a sequence of intensity peaks is formed by spectral interference centred around wavelengths $\lambda_i$. These peaks are delayed from each other at times $t_i$ and experience Self-Phase Modulation (SPM) when being amplified to (sub-)mJ energies. After compression, this results in satellite pulses.}
    \label{fig:intro1}
\end{figure}

This is especially problematic for petawatt systems where parasitic reflections give satellite postpulses up to mJ energies spaced by only a few ps \cite{Apollon-HILAS2024,NIF-2023}. In this case, postpulses are responsible for the generation of prepulses. In turn, this leads to plasma generation or the burning of the sample by the generated prepulse. For burst-mode CPA systems providing picosecond pulse spacings \cite{stummerProgrammableGenerationTerahertz2020a,Burst-OPA,hongtao-2024}, this effect sets an upper burst-energy limit at low pulse numbers $N$. The peak intensity of a chirped few-pulse burst is $N$-times higher than that of a single chirped pulse with the same energy, due to intensity peak formation (see Fig. \ref{fig:intro1}). Therefore, a useful guideline is, that for a single-pulse amplifier that is designed to work with negligible SPM (single-pulse B-Integral maximum $B_{sp,max}<1$) at a given energy, the criterion for SPM-negligible operation in $N$-pulse burst mode is $B_{max}=N\cdot B_{sp,max}<1$. This does not hold for a sufficiently high pulse number $N\gg 1$, where the spacing between the first and last burst pulses becomes comparable to the chirped pulse duration \cite{burst-apl}. Finding solutions to overcome the low-pulse-number limit is motivated because applying only a few pulses allows for the highest pulse energies at a given burst energy.

\begin{figure}[htbp]
    \centering
    \includegraphics[width=\textwidth]{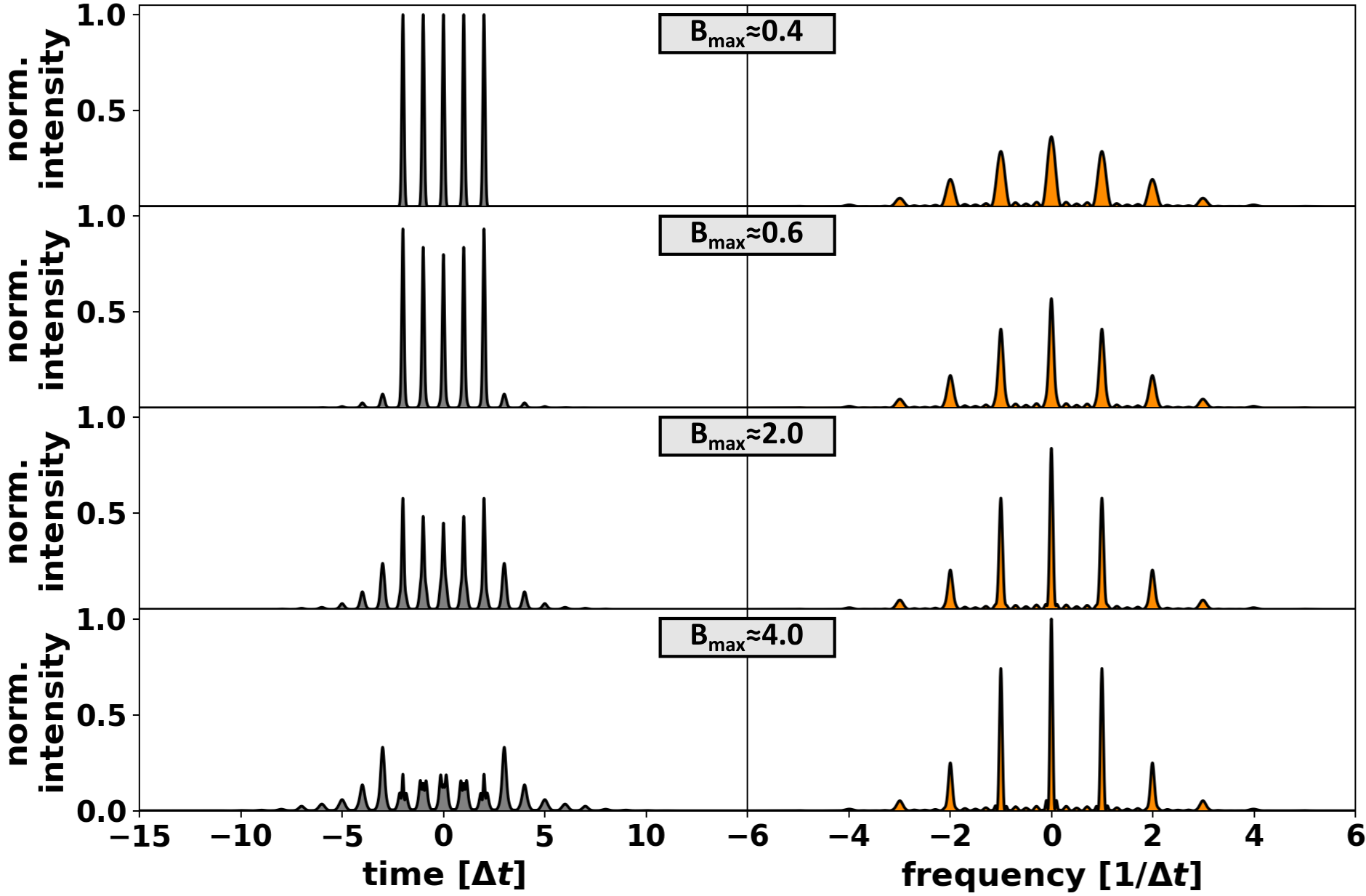}
    \caption{\textbf{Multi-pulse satellite formation with an increasing nonlinearity $N\cdot B_{sp}$ (from top to bottom).}
    Left: Temporal intensity. Right: Spectrum. The chirped pulse duration $\tau_s$ is much larger than the pulse spacing $\Delta t$ ($\tau_s \gg \Delta t$). As the maximum of the total B-Integral approaches 1 ($B_{max}=N\cdot B_{sp,max} \rightarrow 1$), the number and peak intensities of satellites increase, while the spectral peaks become narrower until they start to broaden again due to the action of SPM. Note, that with increasing nonlinearity, there is no spectral broadening unlike in the case of single-pulse SPM.}
    \label{fig:intro2}
\end{figure}

In this paper, we investigate numerically and experimentally the effect of SPM on burst-mode CPA systems up to millijoule (mJ) burst energies, give practical criteria of its relevance, and provide means to suppress its effects. Typically, the presence of nonlinearities is notable in an experiment by investigation of the burst spectrum. In the considered case, however, SPM can lead to a burst-peak narrowing (See Fig. 2). This may easily be overlooked in an experiment with spectrometers providing finite spectral resolution. At the same time, the satellite peak intensities can become quite comparable to the burst pulses, with relative satellite peak intensities of a few tens of percent. In the following, we will elaborate on the SPM effect on a chirped waveform in a time-frequency model based on the Wigner distribution \cite{claasenWignerDistributionTool1980}, which gives a useful qualitative understanding of the underlying process. Experimentally, we generate a pulse pair using our home-built Yb CPA burst-mode system and compare the burst and satellite characteristics with a numerical model. Further, we investigate experimentally the satellite generation in bursts with more than two pulses, and modulate the pulses in their phase, such that suppression of satellite pulses generated by the Kerr effect during amplification is achieved. The underlying results give promising insights that help to deal with nonlinearities arising during the amplification of multi-pulse formats.

\section{Analytical description}
In the following, we give an overview of the analytical description on which the numerical calculation is based on.\\
We assume a stretched burst $E_s^{(B)}(t)$ of $N$ chirped Gaussian pulses $E_s^{(P)}$ that are spaced by a given period $\Delta t$ and have individual carrier-to-envelope (CEP) phases $\phi_n$

\begin{align}
    E_{s}^{(B)}(t) &= \sum_{n=0}^{N-1} E_{s}^{(P)}(t-n\Delta t)\exp{(i\phi_{n})}.
    \label{eq:burstSimple}
\end{align}

For a compressed pulse duration $\tau_c$, a chirped pulse duration $\tau_s$, and a time-bandwidth product $p_0=\tau_c \Delta \omega$, the chirped duration of the burst pulses can be much larger than the compressed duration of the burst $T_B=(N-1)\Delta t$. In the special case of $T_B^2 \ll \tau_c\tau_s/p_0$, the spectral electric field of the compressed burst $\Tilde{E}_c^{(B)}(\omega)$ dictates the time-dependent electric field $E_{s}^{(B)}(t)$ that is realized by the chirping \cite{burst-apl}

\begin{subequations}
    
    \begin{equation}
    E_{s}^{(B)}(t) \propto 
        \exp{\left(i\phi(t)\right)} \Tilde{E}_c^{(B)}(\omega(t))
    \label{eq:burstResult}
    \end{equation}

    \begin{equation}
        \frac{d\phi(t)}{dt} = \omega(t), \hspace{0.25cm} \omega(t) = \omega_0 + \alpha t
    \end{equation}

    \begin{equation}
        \tilde{E}_{c}^{(B)}(\omega) = \tilde{E}_{c}^{(P)}(\omega) \cdot \sum_{n=0}^{N-1}\exp{\left(i(\phi_n-n\omega\Delta t)\right)},
    \label{eq:ComprBurstSpectrum}
    \end{equation}

\end{subequations}

where $\omega_0$ is the center frequency of the pulses, $\alpha$ the chirp rate, $\Tilde{E}^{(P)}_c(\omega)$ the compressed pulse spectral electric field, and $\phi_n$  the CEPs of the burst pulses. In the case of in-phase pulses ($\phi_n=0$), Eq. \ref{eq:ComprBurstSpectrum} gives a burst-typical periodic modulation. Controlling the individual CEPs $\phi_n$ enables the shaping of the burst spectrum, including suppression of the periodic modulation (phase scrambling). For this, it is sufficient to control the relative CEPs within the burst $\phi_{n,0} = \phi_n-\phi_0$, where $\phi_0$ is the first burst pulse's CEP \cite{stummerProgrammableGenerationTerahertz2020a}. 

The B-integral is commonly defined as 

\begin{equation}
    B(t) := \frac{2\pi n_2}{\lambda}\int_0^d |E(z,t)|^2 dz \propto |E_s^{(B)}(t)|^2,
\end{equation}

and is directly proportional to the stretched burst intensity $|E_s^{(B)}(t)|^2$, assuming a homogeneous nonlinear refractive index $n_2$ and a given thickness $d$ of the Kerr medium. \\
The Kerr-modulated burst waveform $E_{mod}^{(B)}(t)$ is calculated to be

\begin{equation}
    E_{mod}^{(B)}(t) = E_{s}^{(B)}(t)\exp{(iB(t)}).
\end{equation}

We multiply the Kerr-modulated burst $E_{mod}^{(B)}(t)$ in the frequency domain with the frequency-dependent compressor response $H(\omega)$ for compression of the burst pulses with a chirp rate $a=1/\alpha$ that corresponds to compression to their initial transform-limited duration

\begin{align}
    H(\omega) &= \exp{\left(-i\frac{a(\omega-\omega_0)^2}{2}\right)}.
    \label{eq:compr}
\end{align}

The resulting compressed Kerr-modulated pulses are not purely Gaussian. For determining the pulse characteristics of peak power $P_{0,n}$, pulse position $t_{0,n}$, and pulse duration $\tau_{0,n}$, fits of the modulated compressed waveform with Gaussian-Lorentzian sums \cite{Gauss-Lorentz-Fit}

\begin{align}
    P_{fit,n}(t) &= P_{0,n} \exp{\left(-4\ln{(2)}(1-m_n)\left(\frac{t-t_{0,n}}{\tau_{0,n}}\right)^2\right)}\cdot\\
    &\frac{1}{1+4m_n\left((\frac{t-t_{0,n}}{\tau_{0,n}})^2+1\right)}
    \label{eq:GL}
\end{align}

led to satisfying results, with Gauss-Lorentz indices $m_n$, for which $m=0$ correspond to a purely Gaussian and $m=1$ to a purely Lorentzian pulse.

\section{Experimental Methods}
For the experiment, we used a mode-locked Yb:KGW oscillator (LightConversion PHAROS, 76 MHz, 1030 nm, 80 fs pulse duration) generating nanojoule pulses that are consequently stretched to 300 ps (FWHM) by a dual-pass transmission grating stretcher in which we applied spectral shaping to precompensate spectral gain narrowing. The burst pulses are picked by an Acousto-Optic Modulator (AOM) where we set the amplitude and phase of the seed pulses. We used the regenerative preamplifier cavity as a time loop in which the round-trip time is slightly different from the oscillator round-trip time for burst pulse accumulation at ps spacings (Vernier Effect)\cite{stummerProgrammableGenerationTerahertz2020a}. For this, the Pockels Cell (PC) voltage is set to a level such that the round-trip losses compensate for the gain during the burst formation before we preamplify the burst to 50 $\upmu$J. In synchronous, we amplified a pulse from the non-AOM-diffracted oscillator train to $\upmu$J energies. The preamplified burst was sent into a cryogenically cooled regenerative booster amplifier for further energy increase. The relevant Kerr media are the BBO crystal in the Pockels cell ($n_{2,BBO}=2.88\cdot 10^{-20}$ W/cm$^2$ \cite{nonlinear-optical-crystals}, $d_{BBO}=40$ mm) and the Yb:CaF\textsubscript{2} laser crystal ($n_{2,CaF_2}=1.9 \cdot 10^{-20}$ W/cm$^2$ \cite{milam1977}, $d_{CaF_2}=10$ mm). The 1/e$^2$ beam diameter was 350 $\upmu\mathrm{m}$ inside the crystal and 850 $\upmu\mathrm{m}$ in the Pockels cell. While satellites generated by the gain saturation are negligible, they can affect the Kerr-induced satellites \cite{didenko2008}. Therefore, we set the number of round trips to 36, which is lower than our booster amplifier's number of round trips for the most efficient energy extraction to avoid saturation effects on the Kerr-induced satellites in this study. In the end, the burst pulses and the reference pulse were compressed by a single spatially multiplexed dual-pass transmission grating compressor. The full laser system is described in detail in \cite{stummerProgrammableGenerationTerahertz2020a,burst-apl}.

\begin{figure}[htbp]
\centering

    \begin{subfigure}[t]{0.4\textwidth}
    \centering
    \includegraphics[width=1.0\linewidth]{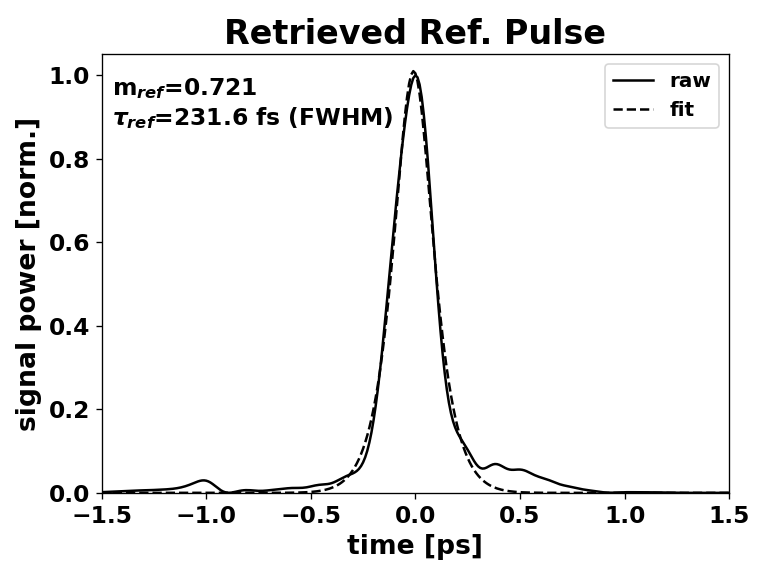}
    \caption{}
    \label{fig:ref}
    \end{subfigure}
    \begin{subfigure}[t]{0.4\textwidth}
    \centering
    \includegraphics[width=1.0\linewidth]{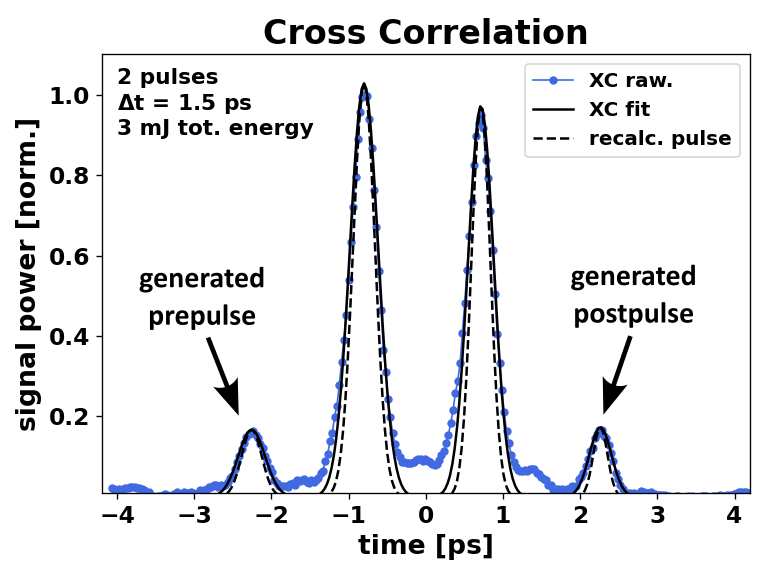}
    \caption{}
    \end{subfigure}

    \caption{\textbf{Temporal characterisation via SHG-FROG and SFG-XC}. a) Reference pulse power retrieved via SHG-FROG. b) SFG-XC measurement of a 2-pulse burst at 3 mJ and 1.5 ps spacing. Blue: SFG-XC raw data. Black Solid: Fit line. Black Dashed: Power of individual pulses derived from the SFG-XC fit under consideration the reference pulse power of a).}

\end{figure}

Experimental characterization was performed by Sum-Frequency Generation cross-correlation (SFG-XC) of the pulse burst with the reference pulse. The reference pulse was slightly longer than the burst pulses because we optimized the gain-narrowing precompensation towards the burst channel. Thus, the individual durations of the burst pulses cannot be determined from the SFG-XC alone. We characterized the single reference pulse by a Second-Harmonic Generation Frequency-Resolved Optical Gating (SHG-FROG) measurement and fitted the time-discrete SFG-XC data with Equ. \ref{eq:GL} to acquire the values of peak powers $P_{0,n}$, pulse positions $t_{0,n}$ and Gauss-Lorentz parameters $m_n$. For determining the correct pulse duration $\tau_{0,n}$, we considered the measured reference pulse duration of 231.6 fs to calculate an accurate value from the SFG-XC fit. This could be done by taking the definition of the cross-correlation

\begin{equation}
    y_{XC,n} = \int_{-\infty}^{+\infty} |E_{n}(t)E_{ref}(t+\tau)|^2d\tau
\end{equation}

with the individual pulse fields $E_n(t)$ and the reference pulse field $E_{ref}(t)$ formulated as Gaussian-Lorentzian sum (Eq. \ref{eq:GL}). For the individual pulse fields, we used the parameters acquired ($P_{0,n}$, $t_{0,n}$, $m_n$), and for the reference field $E_{ref}(t)$, we applied the pulse duration determined by SHG-FROG. Then, we optimized the resulting data $y_{XC}$ towards a fit of the measured SFG-XC with the pulse duration $\tau_{0,n}$ as the optimization variable.

\section{Results}

\subsection{Numerical Results}
\begin{figure}[htbp]
    \centering
    \begin{subfigure}[t]{\textwidth}
    \centering
        \includegraphics[width=\textwidth]{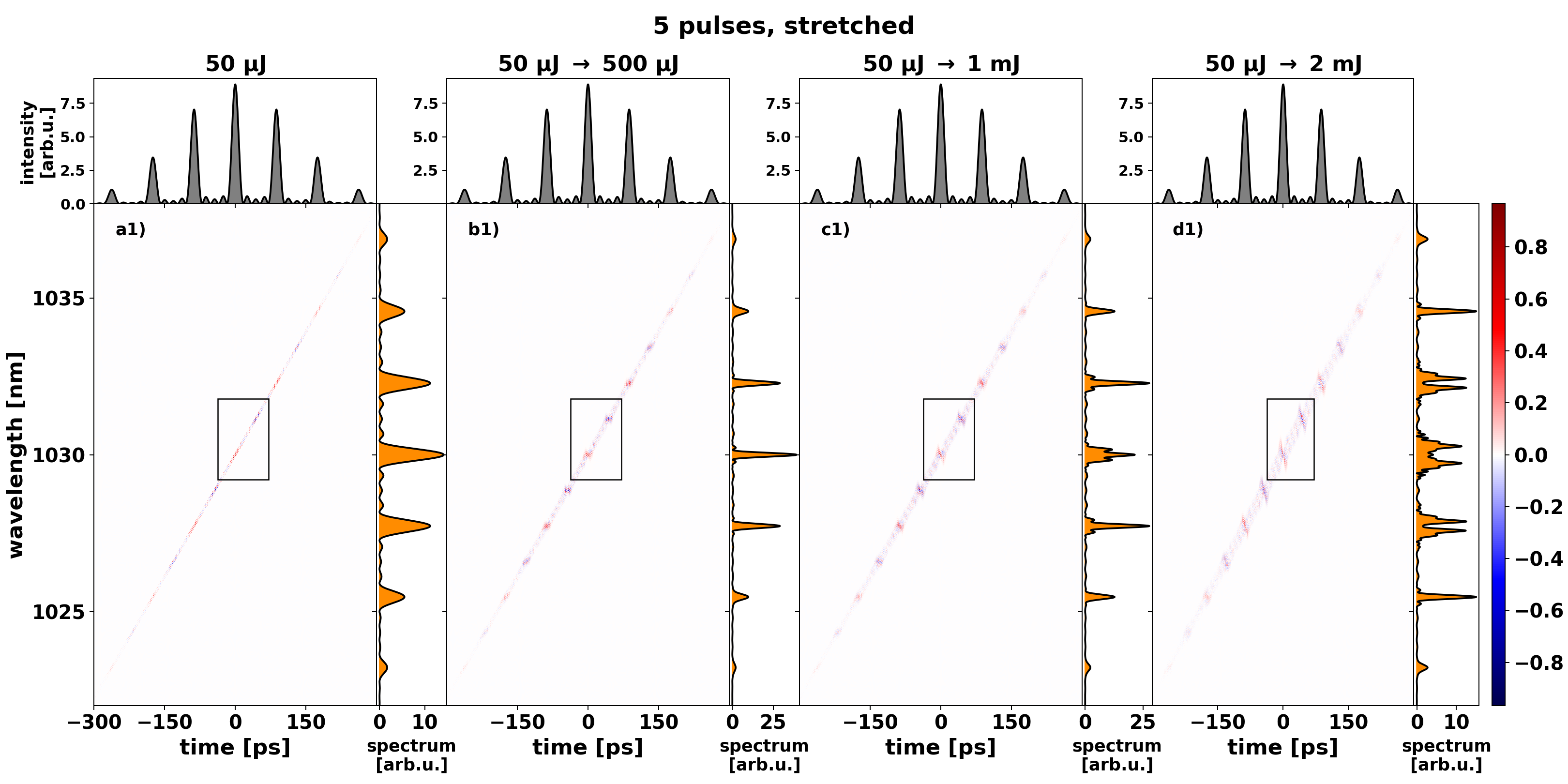}
        \label{fig:wigner_incrE}
    \end{subfigure}

    \begin{subfigure}[t]{\textwidth}
    \centering
        \includegraphics[width=\textwidth]{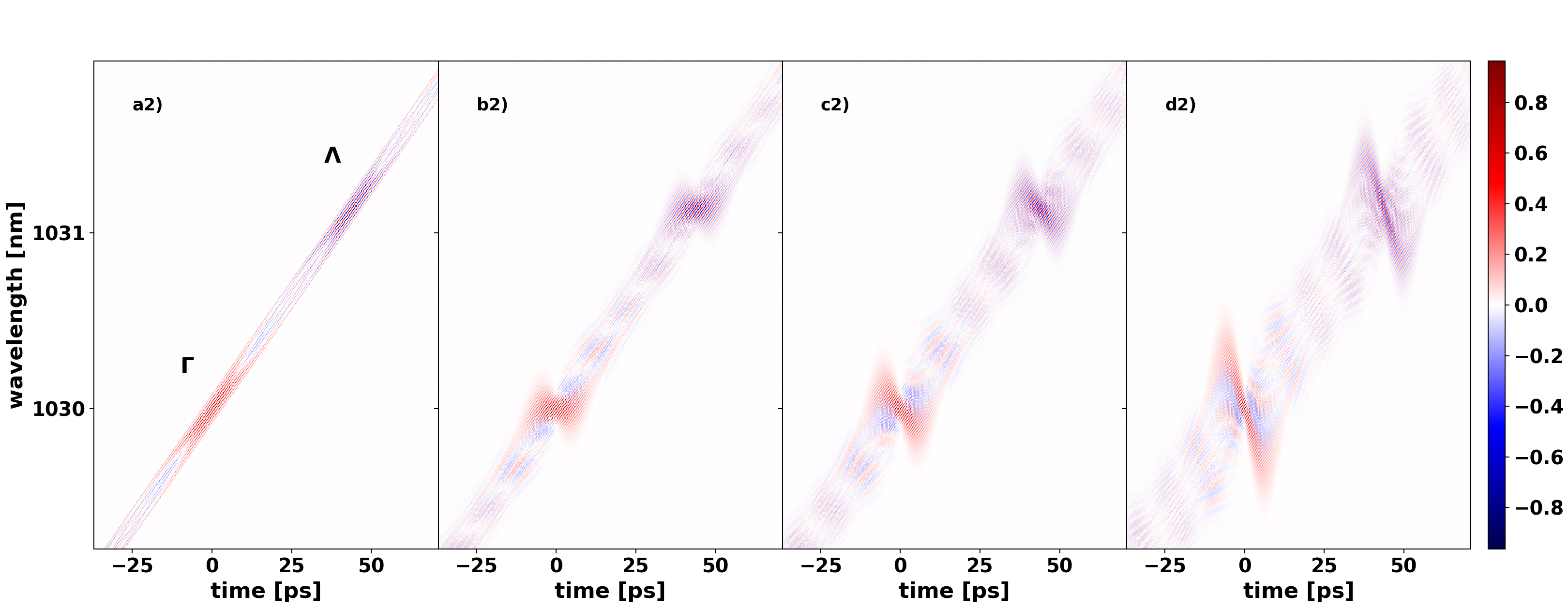}
        \label{fig:wigner_incrEzoom}
    \end{subfigure}
    \caption{\textbf{Wigner distribution of a strongly stretched burst with increasing SPM.} The burst consists of 5 pulses with 1.5 ps spacing, 200 fs TL pulse duration (FWHM) and is chirped to 300 ps (FWHM) a) without SPM. b-d) experiencing SPM while being amplified from 50 $\upmu$J to b) 500 $\upmu$J. c) 1 mJ. d) 2 mJ. The upper plots show the Wigner distribution over a large range, while the lower plots show the corresponding magnification into the area marked with a black square in the upper plot.}
    \label{fig:wigner_incrEFig}
\end{figure}

For a deeper insight into the burst-mode SPM process, we modelled numerically the effect of the B-Integral according to Eqs. \ref{eq:burstSimple}-\ref{eq:compr} and calculated the resulting fields of the stretched SPM-modulated burst $E_{s,mod}^{(B)}(t)$ (see Fig. \ref{fig:wigner_incrEFig}) and the compressed SPM-modulated burst $E_{c,mod}^{(B)}(t)$. With those results, we could determine the corresponding real-valued Wigner distribution $\mathcal{W}(t,\omega)$ according to its common definition \cite{claasenWignerDistributionTool1980}

\begin{equation}
    \mathcal{W}(t,\omega) \coloneqq \int_{-\infty}^{\infty} E(t+s/2)E^*(t-s/2)\exp{(-i\omega s)}ds.
    \label{eq:wigner-def}
\end{equation}

We do so because the time- and frequency-dependent Wigner distribution $\mathcal{W}(t,\omega)$ gives useful insights into the temporal distribution of the frequency components of a complex-valued signal representing the burst electric field \cite{trebinoFrequencyResolvedOpticalGating2000}. It helps in the qualitative understanding of how SPM, being a pure time-dependent phase modulation of the stretched burst, affects the spectrum and the compressed amplified waveform. From the Wigner Distribution, the temporal intensity and the spectrum represent the marginals 

\begin{subequations}
    \begin{equation}
        S(\omega) = \frac{1}{2\sqrt{(\mu_0/\epsilon)}}\int_{-\infty}^{\infty}
        \mathcal{W}(t,\omega)dt
    \end{equation}
    \begin{equation}
        I(t) = \frac{1}{2\sqrt{(\mu_0/\epsilon)}}\int_{-\infty}^{\infty}
        \mathcal{W}(t,\omega)d\omega
    \end{equation}
    \label{eq:wigner-marginals}
\end{subequations}

In Fig. \ref{fig:wigner_incrEFig}, the effect of SPM on the chirped burst is shown for increasing amplified energy. We simulate the amplification in our booster amplifier, where the input burst energy coming from the preamplifier is 50 $\upmu$J at which SPM can be neglected (Fig. \ref{fig:wigner_incrEFig}a). The Wigner distributions show non-zero values located on a diagonal line, resembling a linear distribution of frequencies in time. The spectra shown on the right side determine the temporal waveforms on top due to the linearly chirped pulse duration being sufficiently large compared to the compressed burst duration ($\tau_s \gg T_B^2p_0/\tau_c$). The Wigner distribution reflects this behaviour by featuring regions of purely non-negative values ($\Gamma$ in Fig. \ref{fig:wigner_incrEFig}a2). The much smaller side lobes between the peaks are represented by regions where positive and negative values cancel each other ($\Lambda$ in Fig. \ref{fig:wigner_incrEFig}a2). For this example, we chose a down-chirped waveform as realized in the experiment. In this case, the time-dependent SPM frequency-shift $\Delta \omega (t)$ can counteract the linear chirp, as seen in Fig. \ref{fig:wigner_incrEFig}b. By this frequency shift, we acquire a horizontal distribution of positive-valued components within the Wigner space, leading to sharper spectral lines visible in the spectrum of Fig. \ref{fig:wigner_incrEFig}b1. As SPM becomes stronger due to higher amplification, the spectral peaks broaden comparable to the SPM of individual pulses. We further note, that the effect of SPM on the spectrum is not uniform over the pulse bandwidth, but strongest at the center frequency where the intensity is the highest. Another worthy remark is, that the SPM effect is also visible on the Wigner areas that represent the side lobes, despite the absence of strong temporal intensity at the corresponding temporal coordinates. This indicates an interplay between interference and nonlinear action whose result is the suppression of spectral side lobes in Fig. \ref{fig:wigner_incrEFig}b1,c1.

\begin{figure}[htbp]
    \centering
    \includegraphics[width=\textwidth]{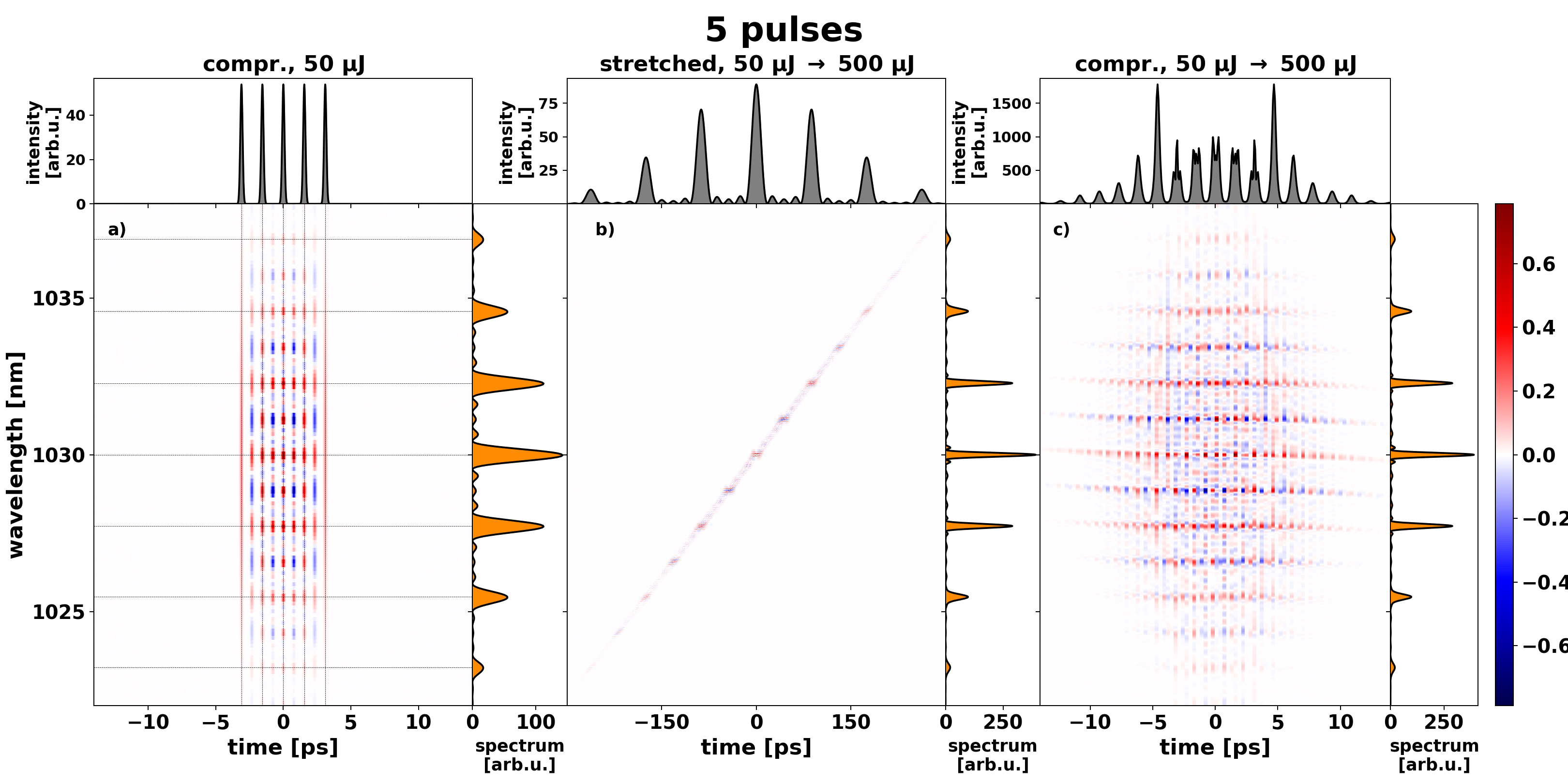}
    \caption{\textbf{Wigner distribution of a strongly stretched 5-pulse burst} a) compressed without SPM. b-c) experiencing SPM while being amplified from 50 $\upmu$J to 500 $\upmu$J with CPA. b) Stretched modulated waveform before recompression. c) Recompressed waveform. 
    1.5 ps pulse spacing, 200 fs TL pulse duration (FWHM). 300 ps chirped pulse duration (FWHM).}
    \label{fig:wigner_compr}
\end{figure}

In Fig. \ref{fig:wigner_compr}, we demonstrate the SPM effect on the compressed burst. First, we explain the Wigner distribution of the non-modulated compressed waveform depicted in Fig. \ref{fig:wigner_compr}a: The interference of each pulse pair located at times $t_i$ and $t_j$ exists in their temporal midpoint $(t_i+t_j)/2$. Thus, the pulse-pair interferences overlap with the individual pulse terms because the burst is periodic. Due to the inversion symmetry of the burst field in time and wavelength, the Wigner distribution results in an inversion-symmetric pattern of the compressed burst without SPM. Horizontal and vertical lines can be drawn along purely non-negative values. Summing up along those lines results in temporal and spectral maxima. In Fig. \ref{fig:wigner_compr}b, we included the plot of the chirped SPM-modulated burst at 500 $\upmu$J, which is the same as in Fig. \ref{fig:wigner_incrEFig}b. Because SPM is strongest around the center wavelength, the compressed Wigner distribution from Fig. \ref{fig:wigner_compr}a is transformed into a distribution that lies within a two-dimensional diamond-shaped envelope (Fig. \ref{fig:wigner_compr}c) and satellite pulses are generated with a lower bandwidth than the burst pulses. This is why the satellite pulses have a longer pulse duration than the original burst pulses. The burst pulses are not recompressed to their transform limit even though they conserve their bandwidth. Neither in our numerical calculations nor in the experiment, we observed the generation of wavelength components outside the individual pulse bandwidth, in contrast to the SPM of ultrashort single pulses.

\subsection{Experimental Results}
We validate our calculation and the characterization method by generating a pair of pulses with equal energies at several above-mJ energies and compare the theoretical and experimental results. Spacings of 1.5 ps and 8.9 ps were chosen to study SPM for small and high temporal separations of pulses.

\begin{figure}[!h]
\centering

    \begin{subfigure}[t]{0.32\textwidth}
    \centering
    \includegraphics[width=1.0\linewidth]{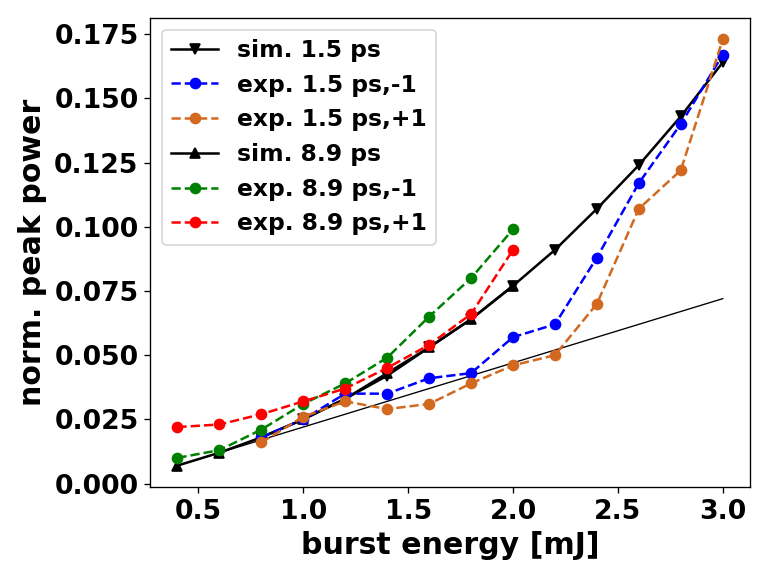}
    \caption{}
    \label{fig:2p-intensities}
    \end{subfigure}
    \begin{subfigure}[t]{0.32\textwidth}
    \centering
    \includegraphics[width=1.0\linewidth]{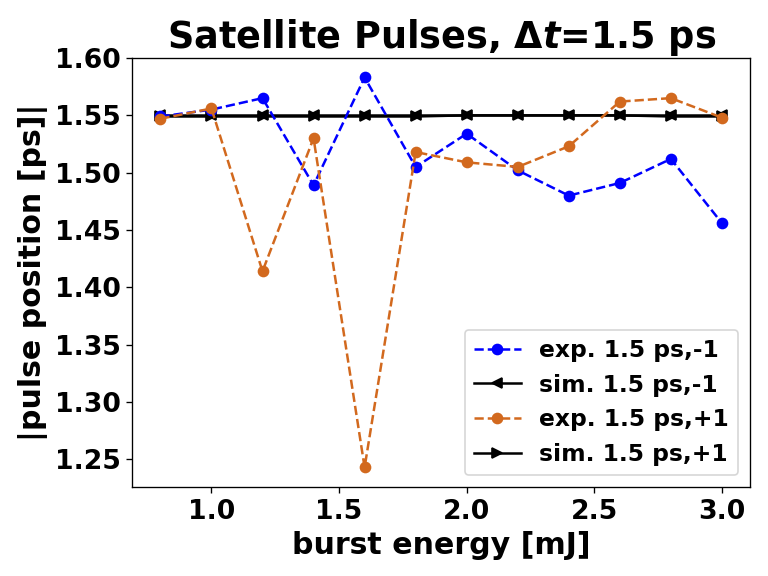}
    \caption{}
    \label{fig:2p-positions_2ps}
    \end{subfigure}
    \begin{subfigure}[t]{0.32\textwidth}
    \centering
    \includegraphics[width=1.0\linewidth]{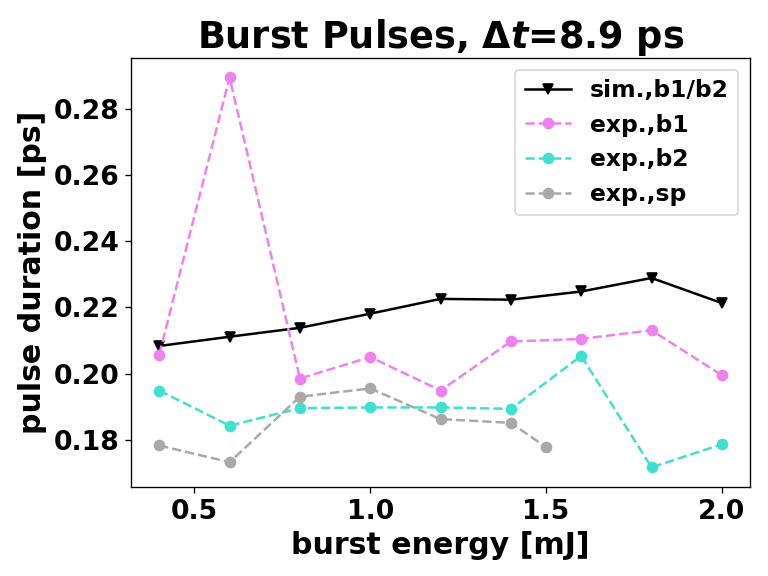}
    \caption{}
    \label{fig:2p-durations-b}
    \end{subfigure}
    
    \begin{subfigure}[t]{0.32\textwidth}
    \centering
    \includegraphics[width=1.0\linewidth]{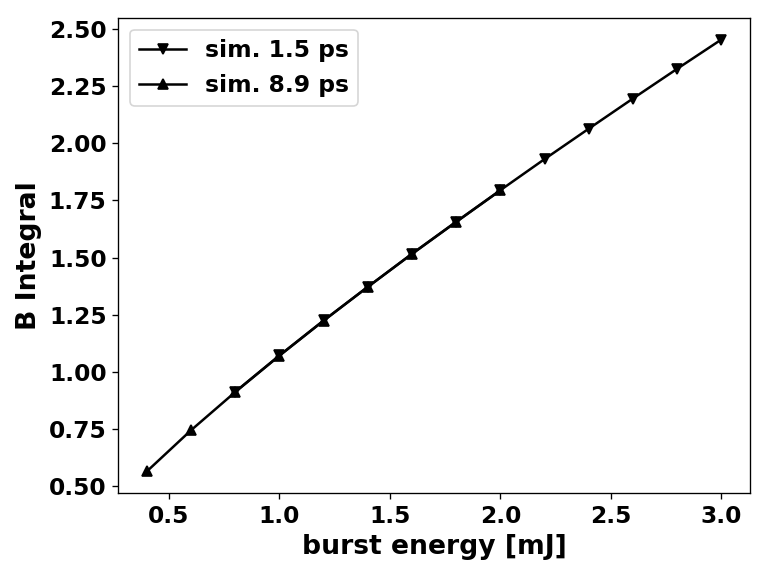}
    \caption{}
    \label{fig:2p-bintegral}
    \end{subfigure}
    \begin{subfigure}[t]{0.32\textwidth}
    \centering
    \includegraphics[width=1.0\linewidth]{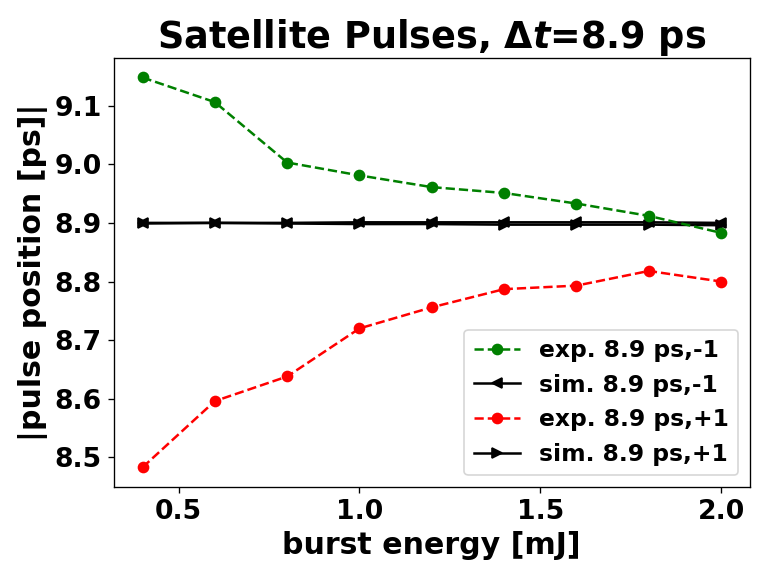}
    \caption{}
    \label{fig:2p-positions_10ps}
    \end{subfigure}
    \begin{subfigure}[t]{0.32\textwidth}
    \centering
    \includegraphics[width=1.0\linewidth]{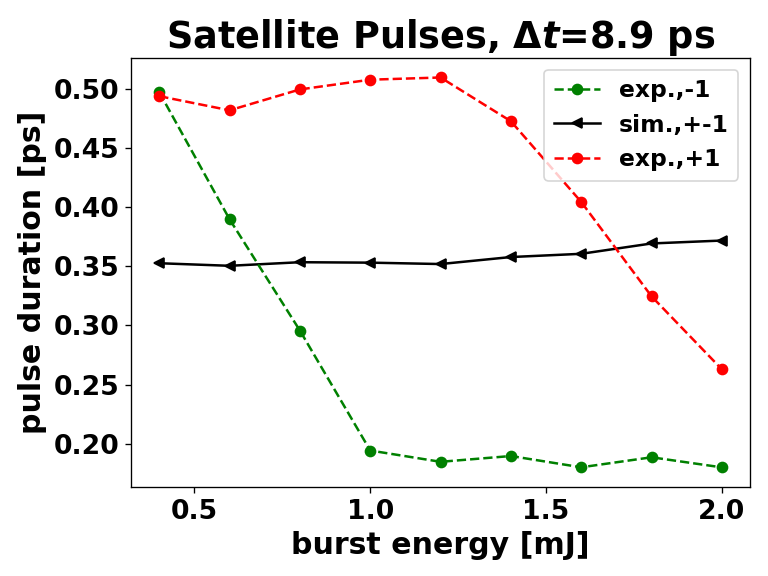}
    \caption{}
    \label{fig:2p-durations_sat}
    \end{subfigure}

    \caption{\textbf{Comparison between simulation and experimental data for burst-mode CPA with 2 pulses.} a) Satellite peak powers vs. output burst energy with pulse spacings of 1.5 ps and 8.9 ps. The satellite peak power is normalized to the average peak power of the burst pulses. Thin Black: Linear approximation from the first two values. b,e) Burst pulse and satellite positions vs. burst energy at a pulse spacing of b) 1.5 ps. e) 8.9 ps. The prepulse/postpulse positions are relative to the first/second burst pulse. c,f) Pulse durations at an 8.9 ps spacing of the c) burst pulses. f) satellites. b) Calculated B integral vs. burst energy. For all subfigures: Black lines correspond to calculated values and coloured lines correspond to measured values. bx: burst pulse x, -1: prepulse, +1: postpulse.}
    \label{fig:2p-stats}
\end{figure}

Fig. \ref{fig:2p-intensities} shows the peak powers of the first-order satellite pulses depending on the output burst energy. We see a good agreement between the calculated curve and the experimental results. For small modulation depths, the $l$-th order satellite peak power depends on the $l$-th power of the B-integral \cite{didenko2008}. With equal pulse energies, the modulation depth is much stronger. Thus, the increase of first-order satellite peak power is nonlinear. SPM does not have substantial effects during single-pulse operation in our amplifier by design ($B_{sp,max} < 1$) at the considered energies. However, the calculated maxima of the burst-mode B-Integral ($B_{max} = N\cdot B_{sp,max}$) were for burst energies higher than 1 mJ consistently above 1 (see Fig. \ref{fig:2p-bintegral}), usually taken as a limit below which nonlinear effects can be neglected. An interesting aspect is the calculated independence of the satellite peak power on the pulse spacing. This is consistent with the $\Delta t$-independent relationship of the burst spectrum $|\Tilde{E}^{(B)}(\omega)|^2$ and the pulse spectrum $|\Tilde{E}^{(P)}(\omega)|^2$ maxima, that is given by

\begin{subequations}

    \begin{equation}
        |\Tilde{E}^{(B)}(\omega)|^2 = \frac{1}{N}|\Tilde{E}^{(P)}(\omega)|^2 \cdot \left(\frac{\sin{\left(\frac{N(\Delta t\omega - \phi_{slip})}{2}\right)}}{\sin{\left(\frac{\Delta t \omega-\phi_{slip}}{2}\right)}}\right)^2
    \end{equation}

    \begin{equation}
        \max{\left[|\Tilde{E}^{(B)}(\omega)|^2\right]} = \max{\left[|\Tilde{E}^{(P)}(\omega)|^2\right]} \cdot N,
    \end{equation}
    
\end{subequations}

under consideration that a pulse carries a $1/N$-fraction of the burst energy. Thus, the independence of nonlinear effects on the pulse spacing indicates that the relationship described in Eq. \ref{eq:burstResult} holds and that the chirped waveform in time is in good approximation given by the burst spectrum because of the dominant linear chirp. Deviations from this behaviour at larger pulse spacings can be attributed to a higher-order chirp. Considering the measured pulse positions (Figs. \ref{fig:2p-positions_2ps} and \ref{fig:2p-positions_10ps}), we observed that the satellite pulse spacings from the burst are close to the fundamental $\Delta t$ spacing. However, in the case of the larger 8.9 ps spacing, the satellite positions depart from the burst pulse spacing $\Delta t$ for smaller energies and approach it for higher energies. This could be given due to the fact, that the Kerr modulation becomes sufficiently pronounced at higher energies. The measured Kerr-modulated burst pulse durations were consistently smaller than the calculated ones (Figs. \ref{fig:2p-durations-b} and \ref{fig:2p-durations_sat}). Further, the calculated satellite pulse durations did not change notably in the considered energy range. However, the measured satellite durations became smaller at larger energies. 

\begin{figure}[!h]
    \centering
    \begin{subfigure}[t]{0.49\textwidth}
    \centering
        \includegraphics[width=\textwidth]{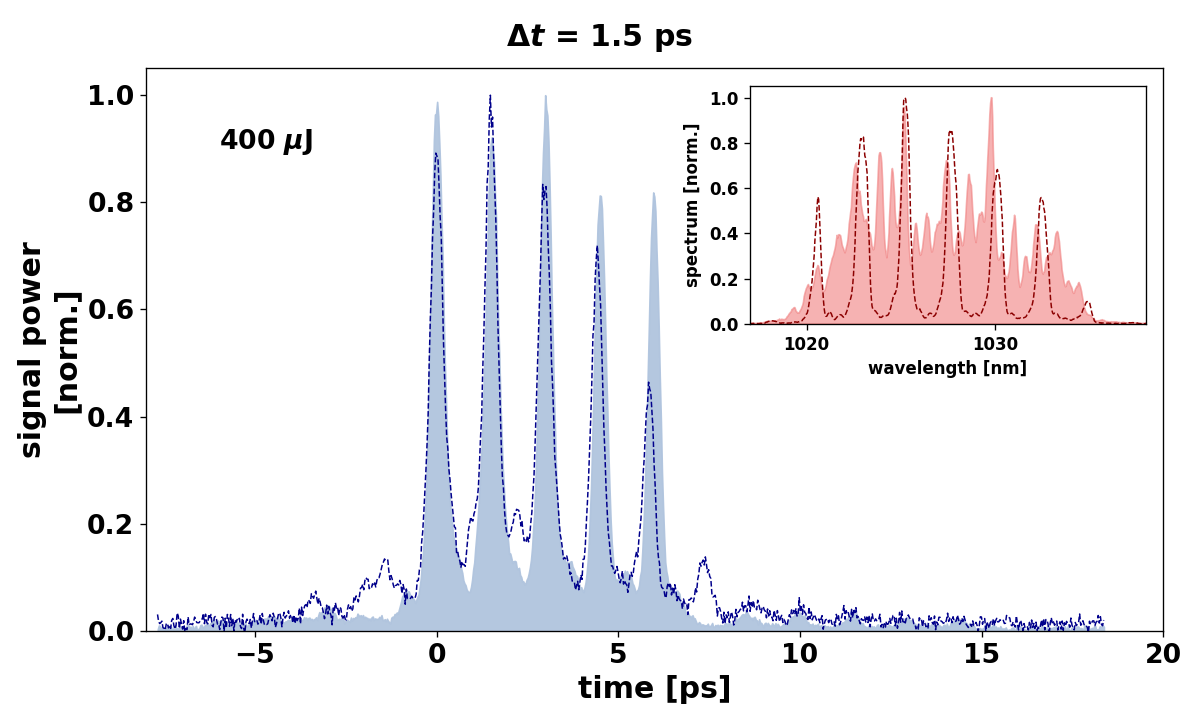}
        \caption{}
        \label{fig:5p-comp_a}
    \end{subfigure}
    \begin{subfigure}[t]{0.49\textwidth}
    \centering
        \includegraphics[width=\textwidth]{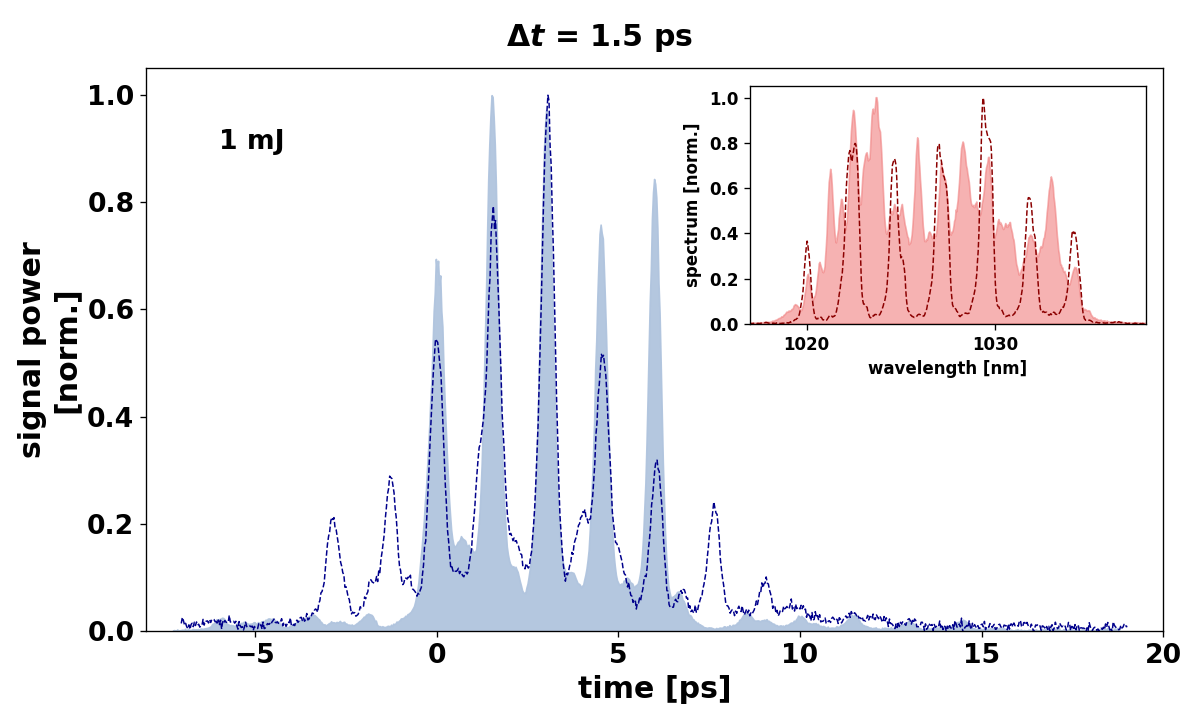}
        \caption{}
        \label{fig:5p-comp_b}
    \end{subfigure}
    \begin{subfigure}[t]{0.49\textwidth}
    \centering
        \includegraphics[width=\textwidth]{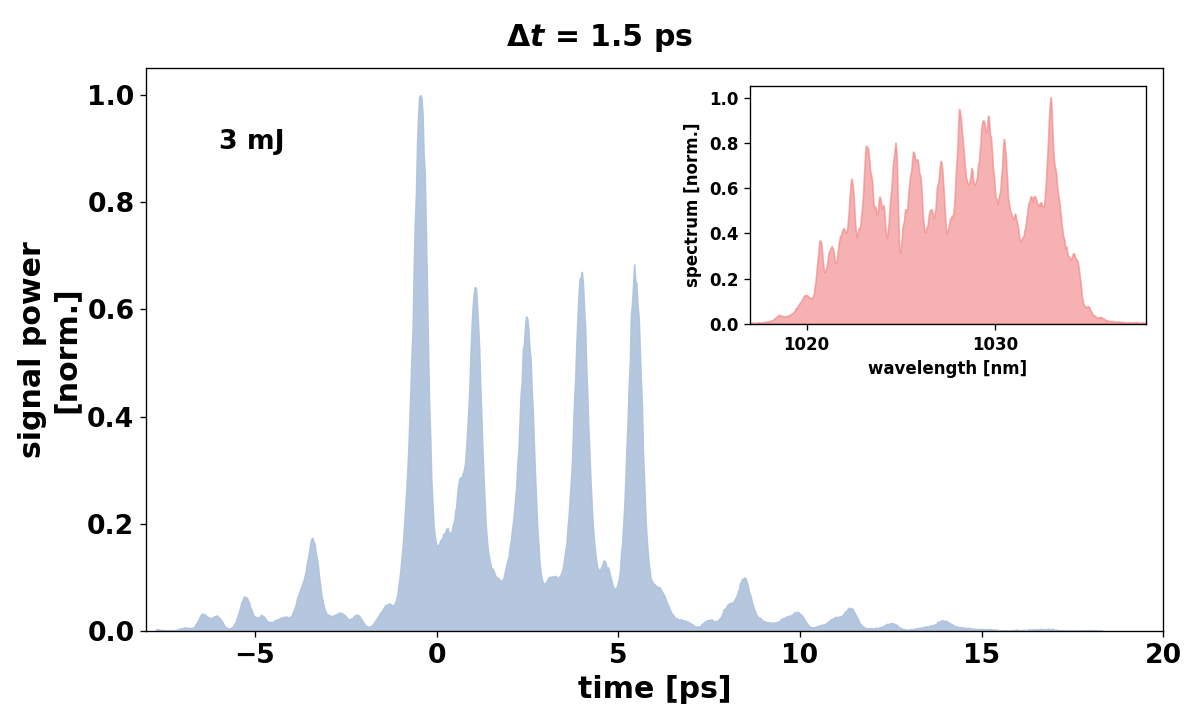}
        \caption{}
        \label{fig:5p-comp_c}
    \end{subfigure}
    \caption{\textbf{Cross-correlations (blue) and spectra (red) of 5-pulse bursts with 1.5 ps spacing and burst energies of a) 400 $\mu$J, b) 1 mJ and c) 3 mJ} by applying no phase modulation (dashed lines) and phase-scrambled modulation (filled areas). At higher energies, we avoid amplification with strong peak formation. Thus, there is no phase-scrambled result in c).}
    \label{fig:5p-comp}
\end{figure}

\begin{figure}[htbp]
    \centering
    \includegraphics[width=0.8\textwidth]{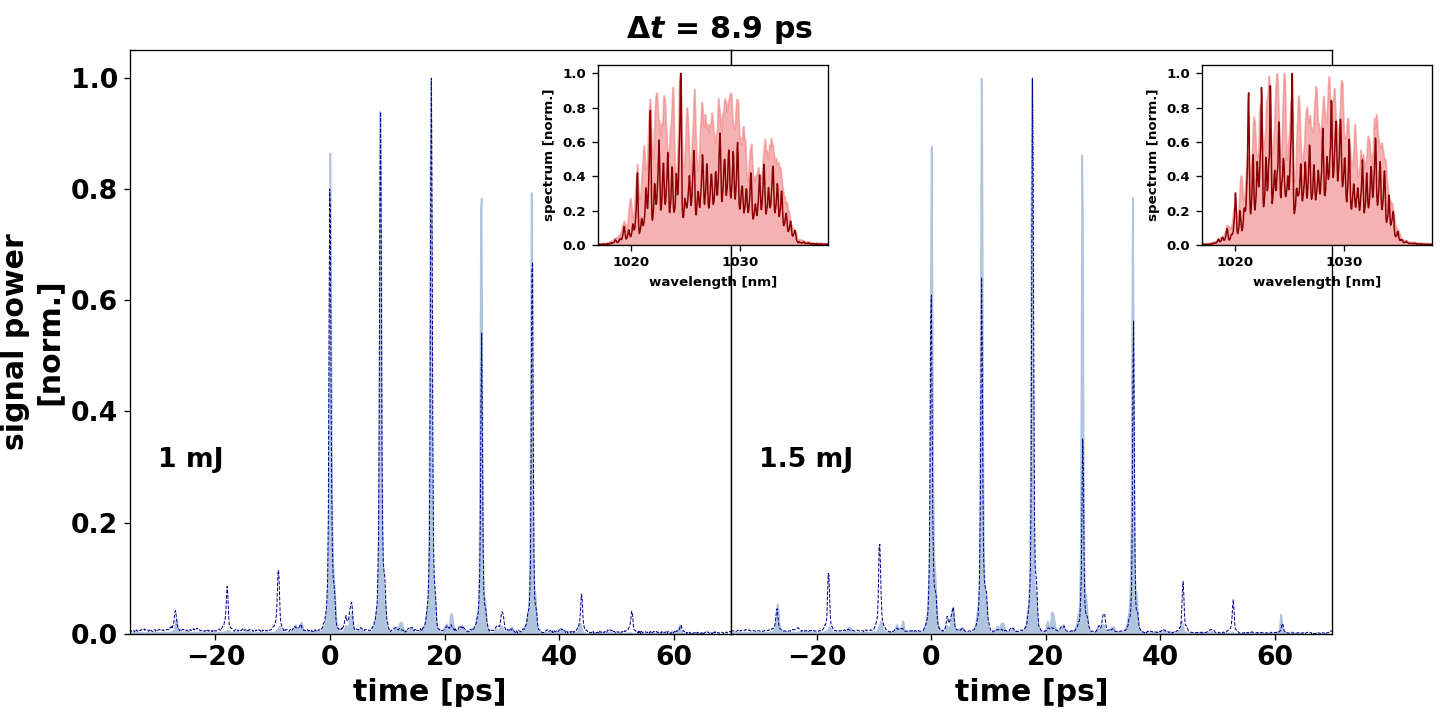}
    \caption{\textbf{Cross-correlations (blue) and spectra (red) of 5-pulse bursts with an 8.9 ps spacing and burst energies of 1 mJ (left) and 1.5 mJ (right)} while applying no phase modulation (dashed lines) and phase-scrambled modulation (filled areas).}
    \label{fig:5p-10ps}
\end{figure}

For more than two burst pulses, it is possible to suppress the spectral periodic modulation by modulating individual pulses in their CEP (phase scrambling). By this, the Kerr-induced satellite generation can be suppressed. Considering relative phases of $0$ or $\pi$, we modulated only the fourth of the 5 pulses by $\pi$ in its phase for the optimum spectral smoothing effect \cite{stummerProgrammableGenerationTerahertz2020a}. SFG-XC measurements were performed together with the measurement of spectra to demonstrate the phase scrambling effect. In Figs. \ref{fig:5p-comp}, we show the results with a burst pulse spacing of 1.5 ps. The satellites started to appear in the SFG-XC measurements when the burst energy was increased to above a few hundred $\upmu$J. The comparison between 400 $\upmu$J and 1 mJ can be seen in Fig. \ref{fig:5p-comp}a-b, where the relative prepulse/postpulse peak power is about 13\%/13\% and 29\%/24\%, respectively, without satellite suppression. However, as soon as we turn on the phase scrambling, visible by a smoothed spectrum, the satellite peak powers could be suppressed to only a few percents. Another notable effect can be seen in Fig. \ref{fig:5p-comp}c, for a 3 mJ burst where the maximum relative satellite peak power was about 17\%: the satellite pulse spacings deviated from the burst pulse spacing because the shaped spectrum did not consist of a single periodic modulation but had a more complex structure. Differences in prepulse and postpulse peak intensities can be generally explained by saturation onset in the gain material at higher energies.
We also observed how the satellite and suppression characteristics change with a larger 8.9 ps pulse spacing (see Fig. \ref{fig:5p-10ps}). For in-phase burst pulses at this spacing, the periodic spectral modulation was already too small for our spectrometer to be resolved. Further, the periodic modulation of the chirped waveform in time is less pronounced because of stronger separated pulses \cite{burst-apl}. At 8.9 ps, the first-order prepulse/postpulse peak intensities were about 11\%/7\% at 1 mJ, in contrast to 29\%/24\% at 1.5 ps spacing at the same burst energy. Prepulse/postpulse peak intensity increased to about 16\%/9\% at 1.5 mJ. When applying the satellite suppression, we see only a few dominant satellites whose relative peak powers are given by a few percents at energies up to 1.5 mJ. 

\begin{figure}[htbp]
    \centering
    \includegraphics[width=\textwidth]{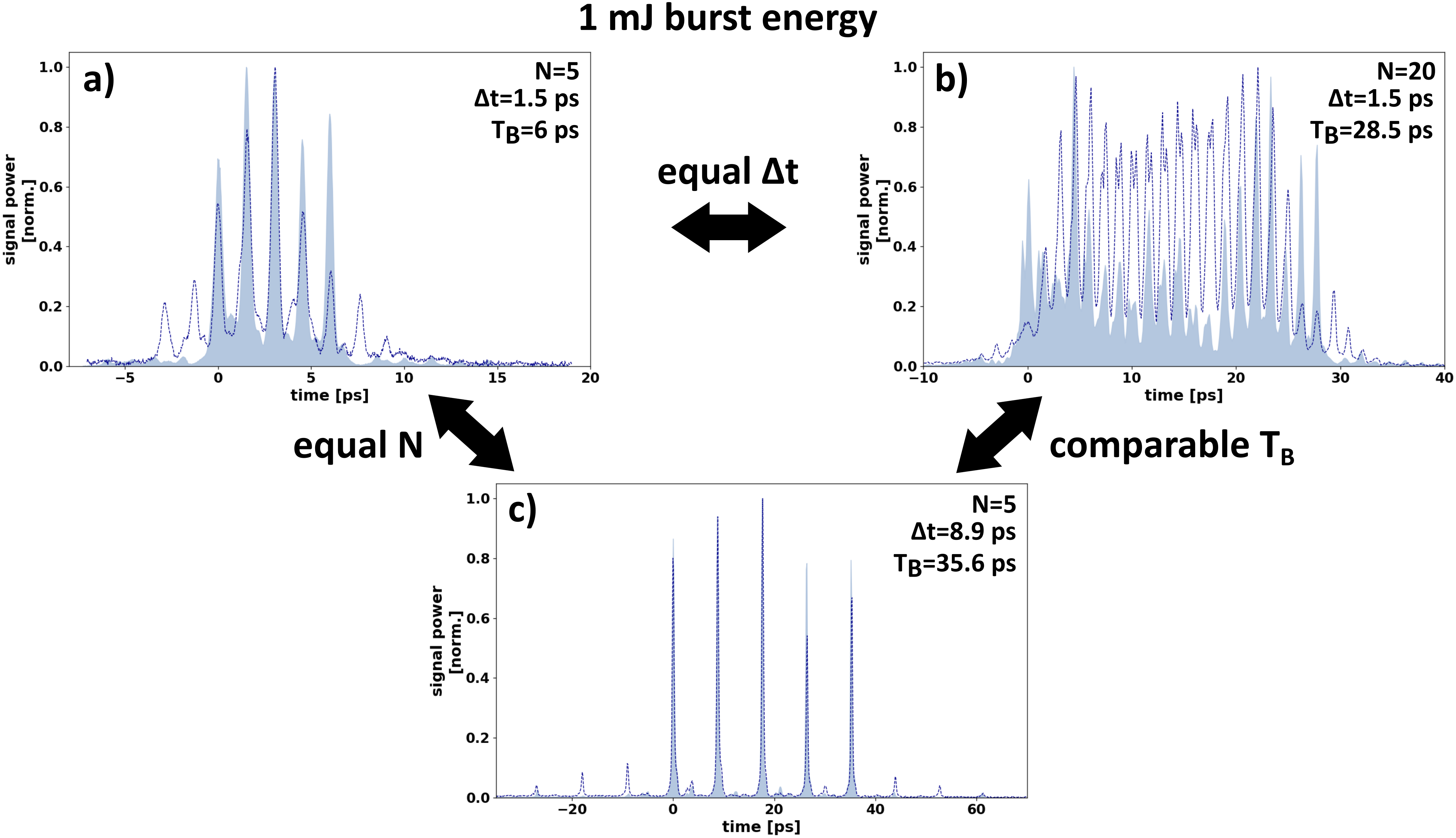}
    \caption{\textbf{Comparison of bursts with pair-wise equal parameters of pulse number $N$, pulse spacing $\Delta t$, and burst duration $T_B$.} Shown are cross-correlations of bursts amplified to a 1 mJ burst energy with a) $N$=5, $\Delta t$=1.5 ps, b) $N$=20, $\Delta t$=1.5 ps, and c) $N$=5, $\Delta t$=8.9 ps, while applying no phase modulation (dashed lines) and phase-scrambled modulation (filled areas).}
    \label{fig:compare_lowN}
\end{figure}

So for smaller pulse spacings at a given pulse number, stronger SPM was observed (Fig. \ref{fig:compare_lowN} a,c). However, in the case of a sufficient increase in pulse number at a given spacing, a decrease in satellite formation is achieved, as can be seen when comparing the temporal intensity profile of bursts consisting of 5 and 20 pulses spaced by 1.5 ps, see Fig. \ref{fig:compare_lowN}a and \ref{fig:compare_lowN}b, respectively. This behaviour reflects the transition of the low-N regime to the high-N regime, which for our system could be identified to start at around 20 pulses at a sub-2 picosecond spacing \cite{burst-apl}. Correspondingly, we refer again to the general rule for linear chirps, that the low-N regime is given if the compressed burst duration $T_B$ is sufficiently small compared to the chirped pulse duration ($T_B^2 \ll \tau_s\tau_c/p_0$). In the low-N regime, pronounced constructive interference of chirped pulses and thus temporal interference peak formation happens. In the high-N regime, the increase in pulse number leads to a smearing of the interference peaks, and thus, to a CPA-typical decrease of the chirp burst waveform. These explanations are further confirmed by the fact, that phase scrambling does not lead to an improved burst waveform in the measured case with 20 pulses (Fig. \ref{fig:compare_lowN}b). Another remark needs to be given to the comparison of Fig. \ref{fig:compare_lowN}b ($N=20,\Delta t=1.5$ ps, $T_B=28.5$ ps), and Fig. \ref{fig:compare_lowN}c ($N=5,\Delta t=8.9$ ps, $T_B=35.6$ ps). Given the comparable burst duration $T_B=(N-1)\Delta t$, both cases are accompanied by a comparable generation of multiple satellite pulses, up to more than $10\%$ of relative peak intensity.

\section{Conclusion}
To conclude, an investigation of SPM effects in multi-pulse CPA up to few-mJ energies has been carried out. We demonstrated the suppression of Kerr-induced satellite pulses by phase modulation of individual pulses and by an increase in burst pulse number. A prepulse/postpulse suppression from 29\%/24\% to only a few percents was achieved in a 5-pulse burst at 1 mJ and a 1.5 ps spacing, and from 16\%/9\% to a few percents at 1.5 mJ and an 8.9 ps spacing by phase scrambling. Further, the satellite generation was less pronounced at larger pulse spacings. This is a promising phenomenon for systems, where prepulses are generated from postpulses coming from parasitic reflections, indicating that the use of parasite-generating components with thicker substrates should further reduce the effect of prepulse generation, due to the smearing effect of the chirped waveform at larger pulse spacings. Another improvement could be given by minimizing the B-Integral by the amplifier design, such as using thinner crystals or increasing amplifier efficiency to reduce the number of round trips. In total, the findings of this work give vital insights into the further development of multi-pulse CPA systems.

\section*{Funding}
Austrian Science Fund (10.55776/F1004, 10.55776/I5590).

\section*{Disclosures}
The authors declare no conflicts of interest.


\bibliographystyle{unsrt}
\bibliography{Satellite-Suppression}






\end{document}